\documentclass{llncs}
\usepackage[utf8]{inputenc}
\usepackage[ruled]{algorithm2e}
\SetAlFnt{\small}
\SetAlCapFnt{\small}
\SetAlCapNameFnt{\small}
\SetAlCapHSkip{0pt}
\IncMargin{-\parindent}

\usepackage{textcomp}
\usepackage{graphicx}
\usepackage{amssymb}
\usepackage{latexsym}
\usepackage{subfig}
\usepackage{listings}
\usepackage{epstopdf}
\usepackage[reqno]{amsmath}
\usepackage{amssymb}
\usepackage{mathtools}
\def\qed{\ifmmode\squareforqed\else{\unskip\nobreak\hfil
    \penalty50\hskip1em\null\nobreak\hfil\squareforqed
    \parfillskip=0pt\finalhyphendemerits=0\endgraf}\fi}

\def\triaforqsd{\hbox{\Large$\triangleleft$}}
\def\qsd{\ifmmode\triaforqsd\else{\unskip\nobreak\hfil
    \penalty50\hskip1em\null\nobreak\hfil\triaforqsd
    \parfillskip=0pt\finalhyphendemerits=0\endgraf}\fi}

\begin{document}

\title{CalcuList: a Functional Language Extended with Imperative  Features }

\author{ Domenico Sacc\`{a} \and Angelo Furfaro}

\institute{DIMES, Universit\`a della Calabria, 87036 Rende, Italy\\
\email{ sacca@unical.it, a.furfaro@unical.it}}

\maketitle

\begin{abstract}

CalcuList (\textit{Calcu}lator with \textit{List} manipulation), is an educational language for teaching functional programming extended with some imperative and side-effect features, which are enabled under explicit request by the programmer. In addition to strings and  lists, the language natively supports json objects.
The language adopts a  Python-like syntax and enables interactive computation sessions with the user through a REPL (Read-Evaluate-Print-Loop) shell. The object code produced by a compilation is a program that will be eventually executed by the CalcuList Virtual Machine (CLVM).
	
%A recent trend in programming languages is a renewed interest towards functional programming, particularly in combining it with other  paradigms, mainly imperative programming. For instance, functional features have been added to existing imperative languages,  e.g., lambda expressions in Java, 
%and new languages, supporting both paradigms, have  been introduced as well. 
%CalcuList (\textit{Calcu}lator with \textit{List} manipulation), is an ``educational'' functional programming language, extended with some side effect features which are enabled under explicit request by the programmer. In addition to lists, the language natively supports JSON objects.
%The language has a  Python-like syntax and it is equipped with a REPL (Read- Evaluate-Print-Loop) shell to establish interactive computation sessions   with the user. During a session, a user  may define a number of global variables and functions and run suitable computations on them as queries, that are computed and displayed on the fly.
%Unlike Python, CalcuList expressions and functions are first compiled and then executed each time a query is issued. CalcuList can be thought of as an ``interactive compiler'' rather than an interpreter and both static and dynamic type-checking are performed. The object code produced by a compilation is a program that will be eventually executed by the CalcuList Virtual Machine (CLVM).

\keywords{Functional Language $\cdot$  Imperative Features $\cdot$ Side Effects } % NOT required for Proceedings

\end{abstract}

%\keywords{Functional Language $\cdot$  Imperative Programming $\cdot$ Side Effects } % NOT required for Proceedings

%\pagenumbering{gobble}
\pagenumbering{arabic}

%\IEEEpeerreviewmaketitle

%\category{F.2}{Theory of Computation }{Analysis of Algorithms and Problem Complexity}
%A category including the fourth, optional field follows...
%\category{H.2.8}{Information Systems}{Database Management}[Database Applications: data mining]
%\category{H.2.2}{Information Systems}{Database Management}{Logical Design}

%\terms{Theory, Design}

%\noindent \textbf{Keywords:} frequent pattern mining.
% ----------------------------------------------------------------
\newsavebox{\leftbx}
\newsavebox{\rightbx}

\section{Introduction}\label{sect:intro}

%The {\em functional programming paradigm}, a form  of {\em declarative programming}, was explicitly created to support a pure functional approach to problem solving. In contrast, most mainstream languages, including object-oriented programming languages such as C++, and Java, were designed to primarily support {\em imperative programming}.
%With an imperative approach, a developer writes code that describes in detail the steps that the computer must take to accomplish the goal, i.e., 
%a program consists of a sequence of commands, which are executed strictly one after the other. In contrast, a functional approach involves composing the problem as a set of functions to be executed: the developer must define carefully the input to each function, and what each function returns.

A recent trend in programming languages is a renewed interest towards functional programming, particularly in combining it with other paradigms, mainly imperative programming.  For instance, functional features have been recently added to existing imperative languages, e.g., lambda expressions in Java~\cite{javalambda},  and new languages, e.g., Scala~\cite{scala}, supporting both paradigms, have been introduced as well. 
During our teaching experiences, we have realized that, 
no matter is the number of functional features introduced in an imperative language, most programmers  continue to use the imperative features and  neglect the functional ones. 

This paper will present a new language, called {\em CalcuList}, that combines functional and imperative programming but does not support
imperative features alone, i.e., any application must have a functional core. Thus, CalcuList essentially belongs to the family of functional languages even though it includes imperative features as well.

Pure functional programming usually designates a specific functional programming paradigm that treats all computation as the evaluation of mathematical functions. Purely functional programming  forbids changing-state and mutable data and mainly consists in ensuring that functions will only depend on their arguments, regardless of any global or local state (i.e., it has no {\em side effects}). Many so-called functional languages are less pure ({\em impure}), as they contain imperative features. For instance, most of the languages of the Lisp Family \cite{Winston_Horn_1986} were designed to be multi-paradigm and mainly present a purely functional interface for many operations (particularly simple mathematics and list/set operations) and an impure interface (more often object-oriented) for things where side effects are desirable, such as I/O. 

Haskell~\cite{Marlow_haskell2010,Hudak:2007} is another important language that is characterized by a purely-functional core. It provides relevant features including polymorphic typing, static type checking, lazy evaluation and higher-order functions. It also has an innovative type system which supports a systematic form of overloading and a module system.

Haskell also expresses side effects but it does it by using a clean approach based on monads that can be thought of as composable computation descriptions \cite{PeytonJones:1993}. The essence of monad is thus separation of composition timeline from the composed computation's execution timeline, as well as the ability of computation to implicitly carry extra data, as pertaining to the computation itself, in addition to its one (hence the name) output, that it will produce when run (or queried, or called upon). This lends monads to supplementing pure calculations with features like I/O, common environment or state, etc.
Thus in Haskell, though it is a purely-functional language, side effects that will be performed by a computation can be dealt with and combined purely at the monad's composition time. 

In contrast with Haskell, CalcuList describes impure effects and actions inside the language itself, although it isolates impure components using an approach, different from Monads, which is derived by the semantic rules of an {\em Attribute Grammar}~\cite{Paakki:1995}. An attribute grammar extends a classical context-free grammar  to specify the context-sensitive aspects of the syntax of a programming language (such as checking that an item has been declared and that its use is consistent with the declaration) as well as its operational semantics (e.g., by defining a translation into lower-level code of a specific machine architecture). In a similar way as a context-free grammar production is enriched with semantic rules to assign values to attributes associated to the symbols occurring in the production, CalcuList enriches the expression defining a function with statements assigning values to global variables and function parameters.

It is interesting to observe that the Curry language~\cite{curry:2011}, defined by extending the functional core of  Haskell, deals with the issue of combining functional and logic programming. 
%The combination of these two paradigms introduces some advantages in terms of both increased declarative power and improved code modularity, reuse, and efficiency of execution. 
CalcuList does not include any logic programming feature, apart from some syntactic notation borrowed from the logic language Prolog~\cite{Clocksin:1984} to represent some basic operations on lists (see Section \ref{sec:list}). 

We stress that CalcuList (\textit{Calcu}lator with \textit{List} manipulation) is an \textit{educational} programming language for teaching the functional paradigm, suitably extended with some imperative features involving side effects. The programmer may use CalcuList as a \textit{pure} functional language, in that side effects are required to be explicitly enabled. The use of side-effects, in some cases, simplifies the developed code or even results in a better computational efficiency.  

CalcuList syntax is rather similar to that of Python~\cite{python} and natively supports processing of strings, lists and json objects~\cite{json}.
The language is strongly typed, but the type checking is mainly dynamic as most of the type checking is done at run-time.
An interactive computation session with the user is established by means of a  REPL (Read-Evaluate-Print-Loop) shell. During a session, a user may define a number of global variables and functions and run suitable computations on them as queries, that are computed and displayed on the fly. CalcuList expressions and functions are first compiled and then executed each time a query is issued. CalcuList can be thought of as an ``interactive compiler'' rather than an interpreter and both static and dynamic type-checking are performed. The object code produced by a compilation is a program that will be eventually executed by the CalcuList Virtual Machine (CLVM).

CLVM is an abstract computing machine that has a set of instructions that are implemented using micro-instructions operating with  three memories (MEM, divided into two parts: STACK and HEAP, CODE and OUTPUT) and a number of registers. The {\em clops} (CalcuList OperationS) of an instruction is the number of micro-instructions executed for its implementation.

%and manipulates three memory areas, each of them organized as an array of 64-bit cells: (1) MEM, which  is divided into two parts: the STACK, which stores global variables and unit frames from the start of the array on, and the HEAP, which stores dynamic data, such as strings, lists and jsons, from the end of the array backward, (2) CODE, storing the instructions of the running CLVM program and (3)  OUTPUT, which is used by the CLVM program during its execution to write the computation results that, after the program termination, are displayed by the CalcuList environment. 
%CLVM is equipped with set of instructions that are implemented using micro-instructions operating with the three memories and the registers. The {\em clops} (CalcuList OperationS) of an instruction is the number of micro-instructions executed for its implementation.
%
The remainder of this paper is organized as follows. Section~\ref{sec:funcCore} introduces basic notions about the CalcuList functional core. Section~\ref{sec:imp} describes the imperative aspects of the language and related side effects.  Finally,  Section~\ref{sec:conc}  draws the conclusions and discusses future work.

% !TeX spellcheck = en_US

\section{An Overview of CalcuList Functional Core}
\label{sec:funcCore}

\subsection{Basic Computations}
The main interface to CalcuList, like other languages (e.g. Python, Scala), is a REPL environment where the user can issue valid expressions (queries in CalcuList), which in turn are parsed, evaluated and printed before the control is given back to the user.

The basic types for CalcuList are five: (1) $double$ (i.e., real number in 64-bit double-precision floating-point format), (2) $int$ (i.e., a 32-bit integer), (3) $char$ (i.e., character represented in the UNICODE format), (4) $bool$ (i.e., a Boolean with values {\tt true} or {\tt false}), (5) $null$ (that has a unique value named {\tt null} as well) and (6) {\tt type} (whose values are all the other types: $double$, $int$, etc.). The language also supports three compound types: $string$ (immutable sequences of characters), $list$ (sequences of elements of any type), and $json$ (JavaScript Object Notation), that is a lightweight data-interchange format.  Details on the usage of strings are given next, whereas lists and jsons will be treated in Section \ref{sec:list} and in Section \ref{sec:json}, respectively. As high order functions are supported, {\tt function} is a type as well.

The first three basic types are {\em numbers} and they are used in arithmetic operations. If the operands are of different type, they are automatically casted to their most general numeric type: $double$ is more general than $int$ that, in turn, is more general than $char$. Moreover, the result of an arithmetic operation of two char terms returns an $int$ value, as it happens in Java.

The arithmetic binary operators that may be used by number terms are {\verb|+|, \verb|-|, \verb|*|, \verb|/|, \verb|//|, \verb|%|}.
The first three operators work just like in most other languages (for example, Java or Python). On the other hand, the operator {\verb|/|} performs floating-point division also when both operands are integer, whereas  {\tt \verb|//|} performs integer division (i.e. quotient without remainder) like in Python. Finally, the modulus operator {\tt \verb|%|} computes the remainder like in Java.  

The language includes unary versions of the operators {\tt +} and {\tt  -}  to change sign to a term. Moreover,  the unary operator {\tt  -} applied to a char term yields an integer result. As we shall see later, the operator {\tt +} is further overloaded as it may be used for concatenating strings and lists. 

The language includes comparison operators {\tt >, >=, <, <=}, equality {\tt ==} and inequality {\tt !=}, which work just like in most other languages (for example, Java or Python). The two operands of a comparison operator must be both: (1) numeric, or (2) $bool$ (with {\tt false < true}) or (3) $string$ (with lexicographic order). On the other hand, the two operands for equality and inequality may be of any (possibly heterogeneous) type.  Finally, the logical operators follows Java syntax: {\tt !} (not), {\tt \&\&} (and), {\tt ||}  (or). Short-circuit evaluation is adopted for binary logical operators, i.e., given {\tt A \&\& B} (resp., {\tt A || B}),  {\tt B} is evaluated only if {\tt A} is true (resp., false).

A string in CalcuList is an immutable sequence (possibly empty) of characters. Two strings can be concatenated by the overloaded operator \verb|+|  returning a new string. Like in Java, it is also possible to concatenate a string with a character (but not the reverse) -- the latter one is automatically casted to a string. Slice operators \`a la Python are available to extract a substring. Given a string {\em s},  $s[i]$ returns the $i$-th character of $s$,
$s[i\mbox{:}]$ returns the substring of $s$ starting from the $i$-th character up to its end; $s[\mbox{:}i]$ returns the substring of $s$ starting from the initial character up to $(i-1)$-th character of $s$; $s[i\mbox{:}j]$ returns the substring of $s$ that begins at the index $i$ and ends up to the index $j-1$.

The syntax for an expression is like in Java. An expression $e$ is either {\em simple} or {\em conditional} with format $e_1$ \verb|?| $e_2$ \verb|:| $e_3$, where $e_1$ is a logical (i.e., with type $bool$) expression and $e_2$ and $e_3$ are expressions (possibly conditional in their turn). The meaning of a conditional expression is: its value is equal to the value of $e_2$ if $e_1$ evaluates to $true$ or, otherwise, to the value of $e_3$.

The language allows the user to define global variables. A {\em global variable} $V$ is defined as ``$V = e$'', where $e$ is an expression: the type of $V$ is inferred from the type of $e$ and its value is kept until a later re-definition, which may change also the type.
Compound assignment operators {\tt +=, -=, *=, /=} are available as well.

At the start of a CalcuList session, the set of global variables is empty. The {\em state} of a session is given by the set of all defined global variables with the most recently assigned values and types for them.

A  {\em query} is an expression $e$ prefixed by the \verb|^| symbol and prints the value of the expression at the current state. Examples of variable definitions and queries are shown in the CalcuList session of Figure~\ref{fig:prel} - they are prompted by ``\verb|>>|'' and are terminated by a semicolon. The state at point of the session indicated by the comment ``/*  checkpoint 1 */'', is: $\{  \mbox{$<$}\mbox{x}, \,  bool, \tt{true}\mbox{$>$}$, $\mbox{$<$}\mbox{y},  string, \tt{\verb|"|Hello \; World\verb|"|}\mbox{$>$} \}$. 

In addition to queries, the user may submit ``service commands'' (preceded by \verb|!|) to inquiry about the variables and functions defined during the session, to list the history of all definitions, to print the state of the internal memory, to display the number of micro-instructions performed in the last execution, to enable or disable the debugger and the tail recursion optimizer, to save or import a session of definitions and so on. 

\vspace{-2mm}

		\lstset{upquote=true}
\begin{lrbox}{\leftbx}%
	\begin{minipage}{.89\linewidth}%
		\begin{lstlisting}
>> x=2+.1;    >> x *= 2;     >> ^x;
4.2
>>x='A'+1=='B';    >> ^x;
true
>> y="Hello "+"Worl"+'d';   >>^y;
Hello World
>> ^y[0:1]+'i'+y[5:];
Hi World
>>/* checkpoint 1 */
>>fibe1(x,f2,f1,k) :  x==k?  f1: fibe1(x,f1,f1+f2,k+1); 
>>fibe(x) :  x <= 1?  x: fibe1(x,0,1,1);
>>z=fibe(10);
>>^z
55
>>^z@type
int

       \end{lstlisting}
\end{minipage}%
\end{lrbox}

%\vspace{-0,5cm}
\begin{figure}
\framebox[1.1\width] {
\usebox{\leftbx}
}  
\caption{\em Basic Computations}
\label{fig:prel} 
\vspace{-0,7cm}
\end{figure}

A function is defined by giving it a name, followed by: its comma-separated parameter names (included in parentheses), the colon symbol ``\verb|:|'' and the (typically conditional) expression that computes the value of the function. Parameters as well as the return value of a function are defined without specifying types for them, so that  static type checking is performed only for constant operands. A function is compiled while it is defined and, as for Python, the type checking  is done at run time, after its call, when all the types become available. In Section \ref{sec:conc} we shall discuss a forthcoming extension of CalcuList for inferring types for some function parameters. Lazy evaluation is not supported and at the moment, there is no plan to introduce it in the future.

No side effects are allowed in the basic definitions of functions, i.e., functional operations do not modify current global variables and always create new data objects. Differently from Python, global variables are not accessible inside a function so that a possible change of state does not affect the function behavior. An example of function computing a Fibonacci number by a forward execution, starting from the first two Fibonacci numbers, is shown in Figure~\ref{fig:prel}. 
The value 55 of the fibonacci number 20 is computed and stored into the variable {\tt z}: this value is of type $int$ as confirmed by the answer to the query {\tt \verb|^|z@type}.

Observe that, given an expression {\tt e}, whose value is one of the eight admissible types, the casting operator {\tt e@type}  returns the type of  {\tt e} -- in the example of Figure~\ref{fig:prel}, {\tt \verb|^|z@type} returns the type $int$ of the value 55. The eight types can be also directly assigned as values to variables and their type is $type$, i.e., {\tt \verb|^|int@type} returns $type$. 
The state at the end of the session is: $\{  \mbox{$<$}\mbox{x}, \,  bool, \tt{true}\mbox{$>$}$, $\mbox{$<$}\mbox{y},  string, \tt{\verb|"|Hello \; World\verb|"|}\mbox{$>$}$,  $\mbox{$<$}\mbox{z},  int, 55\mbox{$>$}\}$. 

\subsection{Lists}\label{sec:list}

The {\em list} is a powerful compound data type used by CalcuList for constructing dynamic data structures and implementing recursive algorithms.  A list {\tt L} consists of a number (possibly zero) of comma-separated elements between square brackets. As in Python, the elements can be of any type, including list and json, and heterogeneous.  
They 
are numbered with an index starting from zero: the first element (called the {\em head} of the list) is {\tt L[0]} (also denoted simply by {\tt L[.]}), the second element is {\tt L[1]} and so on. List assignment is a shallow operator as the list elements are shared by the two lists involved in the assignment. Two lists can be compared only with the operators {\tt ==} and {\tt !=}. 

A list can be extended by adding additional comma-separated elements on top, followed by the append operator (the bar ``\verb+|+'') - the syntax has been inspired by Prolog~\cite{Clocksin:1984}. For instance, given a list {\tt L}, {\tt M=[x,y\verb+|+L]} extends {\tt L} by adding two new elements, {\tt x} and {\tt y}, on top of {\tt L}. Thus, append is a shallow operator so that, for instance, if an element of {\tt L} is changed, the corresponding element in {\tt M} is changed as well. Also the concatenation {\tt L1+L2} of two lists is shallow as the elements of {\tt L2} are afterwards shared by {\tt L1}. 

Another shallow operator on a list {\tt L} is \verb|L[>]|, which returns the (possibly empty) {\em tail} of {\tt L}, i.e., the list starting from the element {\tt L[1]}. If {\tt L} contains exactly one element then \verb|L[>]| is the empty list {\tt []}.  On the other hand, if {\tt L} is empty, then the evaluation of \verb|L[>]| will report an execution error. 

Slice operators on a list {\tt L} (say with $n$ elements) are deep in the sense that they clone the elements. In particular, {\tt  L[:]} clones the whole list {\tt L} and
\verb+L[i:]+ clones the sublist of {\tt L} from the element {\tt L[i]} to the end; in the same way, ${\tt L[i_1\verb|:|i_2 ]}$ and {\tt L[\verb|:|i]} clone the corresponding sublists. 

\vspace{-1.5mm}

\begin{lrbox}{\leftbx}%
	\begin{minipage}{.89\linewidth}%
		\begin{lstlisting}
>>member(x,L): L !=[] && (x==L[.] || member(x,L[>]));
>>listRev(L): L==[]? []: listRev(L[>])+[L[.]];
>>rev1(L,R): L==[]? R: rev1(L[>],[L[.]|R]);
>>rev(L): rev1(L,[]);
>>range(x1,x2): x1>x2? []: [x1|range(x1+1,x2)];
>>L1= range(1,1000); L2 = range(1,2000);
>>^listRev(L1);
[1000, 999, 998, 997, ... ]
>> !clops;
2795726
>>^listRev(L2);
[2000, 1999, 1998, 1997, ... ]
>> !clops;
10591226
>>^rev(L1);
[1000, 999, 998, 997, ... ]
>> !clops;
295308
>>^rev(L2);
[2000, 1999, 1998, 1997, ... ]
>> !clops;
590308
>>merge(O1,O2): O1==[]? O2[:]:O2==[]? O1[:]: O1[.]<O2[.]? 
                [O1[.]|merge(O1[>], O2)]:  [O2[.]|merge(O1,O2[>])];



       \end{lstlisting}
\end{minipage}%
\end{lrbox}

%\vspace{-0,3cm}
\begin{figure}
\framebox[1.1\width] {
\usebox{\leftbx}
}  
\caption{\em Functions on Lists}
\label{fig:list} 
\vspace{-0,7cm}
\end{figure}

Figure~\ref{fig:list} includes a number of functions for handling lists. The first one is the classical {\tt member} function that checks wether an element belongs to a list. The  {\tt listRev} function  constructs the reverse of a given list. At the generic recursion step on a current list {\tt L}, the function concatenates the tail of {\tt L}, recursively computed, with its head. The complexity if {\tt listRev} is quadratic in the size of the list length, say $n$ - actually the number of operation is $n (n-1) / 2$. 
A linear-time implementation is given by  {\tt rev}, which uses ``tail recursion'', which enables avoiding to scan the entire first term list in the concatenation operation -- note that in CalcuList an element of a list cannot be directly accessed by its index  but only after scanning all the previous elements.
To compare the two implementations, we construct  two lists L1 and L2 with respectively 1000 and 2000 elements, by means of the function {\tt range}  and issue the service command
 {\tt !clops} to get the number of micro-instructions (CalcuList micro OperationS) of CalcuList Virtual Machine that have been performed in the last execution. In the example of Figure~\ref{fig:list}, \verb|^|{\tt listRev(L1)} and \verb|^|{\tt rev(L1)} are executed by $k^1_1=2,795,726$ clops and $k^2_1=295,308$ clops, respectively. The number of clops executed by  \verb|^|{\tt listRev(L2)} and by \verb|^|{\tt rev(L2)} are respectively $k^1_2=10,591,226$  and $k^2_2=590,308$. As $k^1_2=3.788 \times  k^1_1$ and $k^2_2=1,999 \times  k^2_1$, such results confirm that 
{\tt listRev} runs in almost quadratic time whereas {\tt rev} runs in linear time.
 
 Let us now raise an important issue about possible side effects that could be induced by the usage of shallow list operations in a function. For instance, 
 the function {\tt listRev}  contains the shallow operation {\tt listRev(L[>])+[L[.]]}. However it is easy to see that the operation does not have any side effect on the parameter {\tt L} since the list corresponding to {\tt listRev(L[>])} is constructed from an empty list (at the last step of the recursion) that is later extended with elements copied from {\tt L} at each previous step. Also
 the  function {\tt rev1} contains a shallow operation {\tt [L[.]|R]}. In this case, the second parameter {\tt R} of the function is modified by the computation. But this parameter has been passed by the function {\tt rev} with the initial value of an empty list so that no side effects will affect the session state. At the moment, the side effect check is left to the programmer. As we shall discuss in Section \ref{sec:conc}, automatic checks will be introduced in a forthcoming version of CalcuList. 
 
 Figure~\ref{fig:list} includes another function that makes use of shallow operations: {\tt merge(O1,O2)}. In this case, the exit conditions do not return an empty list as for {\tt listRev} but residual portions of the two parameters {\tt O1} and of {\tt O2} that are, however, cloned to avoid possible side effects.
 
\subsection{JSON}
\label{sec:json}

Json         (JavaScript Object Notation)~\cite{json} is language-independent data interchange format, which has been originally introduced as a subset of  the JavaScript scripting language. 
%Because it is a very comfortable notation, being both human readable and easy to parse and process, 
The use of json is now widespread in many applications, e.g. web-services,  and support for it has been added to the standard libraries of many programming languages.  Recently, a great deal of interest is focused on {\em json database stores} that are NoSQL database systems using json documents instead of relation tuples \cite{NoSQL}.

CalcuList natively supports json objects (referred to simply as $json$), which are at runtime represented as  (possibly empty) sequences of fields separated by comma and enclosed into curly braces. A field is a pair (key, value) separated by a colon: $key$ is a string and $value$ can be of any type: double, int, char, bool, string, null, list and (recursively) json. Two fields of a json object must not have the same key.

In Figure \ref{fig:json} we define a variable { \tt emps} as a list of three jsons, each representing an employee. For instance, {\tt emps[1]} is the employee with name {\tt "e2"}, aged 32 and working for the projects {\tt "p1"} and {\tt "p2"}.
Given a json $J$ and a key $k$, $J[k]$ denotes the value of the field of $J$ with key equal to $k$ -- if there is no such a field then the value $null$ is returned as it happens for {\tt \verb|^|emps[0]["projects"]} (the string \verb|%*| 
next to the query is a display option to force the writing of the null value). Given a value $v$, $J[k]=v$ modifies the value for the field if $J$ includes the key $k$ or otherwise, $J$ is extended with a new field with key $k$ and value $v$. The assignment of a json to a variable is a shallow operation. To clone a json $J$, it is sufficient to type $J[:]$.
Two jsons can be compared only with the operators {\tt ==} and {\tt !=} and the equality is true if and only if the two json values point to the same data structure in the memory. The length of a json $J$, i.e., the number of its fields, can be obtained by invoking the built-in function {\tt \_len(J)}.

\begin{lrbox}{\leftbx}%
	\begin{minipage}{.89\linewidth}%
		\begin{lstlisting}
>>emps = [ { "name": "e1", "age": 30 },
           { "name": "e2", "age": 32, "projects": [ "p1", "p2" ] },
           { "name": "e3", "age": 28, "projects": [ "p1", "p3" ] } ];
>>^emps[2]["projects"];
[ "p1", "p3" ] 
>>^emps[0]["projects"] %*;
null
\end{lstlisting}
\end{minipage}%
\end{lrbox}

\vspace{-0,2cm}
\begin{figure}[t]
\framebox[1.1\width] {
\usebox{\leftbx}
}  
\caption{\em Json Definition and Manipulation in CalcuList}
\label{fig:json} 
%\vspace{-0,7cm}
\end{figure}

\subsection{Higher-Order Functions}

CalcuList supports higher-order functions since a function parameter can also be a function and a function may  return a function. 
A function parameter {\tt f} is written as {\tt f/}{\em n}, where $n$ is the arity of the function {\tt f}. In addition, adding {\tt /}{\em n} after a function head {\tt g(\dots)} prescribes that the function {\tt g} must return a function with arity $n$. As an example, consider the higher-order function {\tt twice} that takes a function, and applies the function to some value twice -- note that lambda functions are written in Python like syntax: 
%\\ \centerline{ {\tt twice(f/1)/1: lambda x: f(f(x));}} \\ 
\vspace{-0.12cm}
\begin{center}
{\tt twice(f/1)/1: lambda x: f(f(x));}
\end{center}
\vspace{-0.12cm}
 \noindent
For instance, the query {\tt \verb|^|twice(lambda x: x+3)(7)} returns the value 13.

CalcuList's syntax for denoting function arity has two advantages: function terms can be easily singled out and an early static check on their arity can be easily done, whereas a run-time check would require a more complex framework.  

Some examples of higher-order function are presented in Figure~\ref{fig:highOrder}.
The function {\tt map} receives a list {\tt L} and two functions as parameters, both with arity 1: the function {\tt f} checks whether an element has a certain property and, in case the test succeeds,  {\tt m} maps the element into another value.  The function {\tt map} scans all elements of {\tt L}, for each of them it performs the possible mapping and eventually returns the list of the results. The first map query  filters all numbers divisible by 2 or by 3 (see function {\tt d2or3}) in the range from 1 to 10 and replaces each of these values with their cube (see the lambda function) -- let $M$ denote the result of this query. 

\begin{lrbox}{\leftbx}%
	\begin{minipage}{.89\linewidth}%
		\begin{lstlisting}
>>map(L,f/1,m/1) : L==[]?[]: f(L[.])? [m(L[.])|map(L[>],f,m)]:  map(L[>],f,m);
>>d2or3(x) : x%2==0 || x%3==0;
>>^map(range(1,10),d2or3,lambda x: x*x*x);
[ 8, 27, 64, 216, 512, 729, 1000 ]
>>reduce(L,f/2,init) : L==[]? init: f(L[.],red(L[>],f,init));
>>sum(x,y):x+y; >>prod(x,y):x*y;
>>^reduce(map(range(1,10),d2or3, lambda x: x*x*x),sum,0);
2556
>>^reduce(map(range(1,10),lambda x:x%2==0&&x%3==0,lambda x:x*x),prod,1);
36
>>jsFilter(LJ,filtC/3,K,V): LJ==[]?[]:filtC(LJ[.],K,V)? 
    [LJ[.]|jsFilter(LJ[>],filtC,K,V)]: 
    jsFilter(LJ[>],filtC,K,V); 
>>select1KV(J,K,V): J[K]!=null && J[K]==V; 
>>^jsFilter(emps,select1KV,"age",28);
[ {"name": "e3",  "age": 28, "projects": ["p1", "p3"] } ]
>>select1KinV(J,K,V) : J[K]!=null && member(V,J[K]); 
>>^jsFilter(emps,select1KinV,"projects","p1") ;
[  { "name": "e2", "age": 32, "projects": [ "p1", "p2" ] },
   { "name": "e3", "age": 28, "projects": [ "p1", "p3" ] } ]
       \end{lstlisting}
\end{minipage}%
\end{lrbox}

%\vspace{-0,4cm}
\begin{figure}
\framebox[1.1\width] {
\usebox{\leftbx}
}  
\caption{\em Higher Order Functions on Lists and Jsons}
\label{fig:highOrder} 
\vspace{-0,7cm}
%\vspace{-0,3cm}
\end{figure}
             
The subsequent function {\tt reduce} receives a list {\tt L}, a function parameter {\tt f} (mapping two values into another value) and an initial value that is returned at final stage of the recursion, when {\tt L} reduces to an empty list. Then {\tt reduce} applies  {\tt f}  to two actual parameters: the list head and the result of {\tt reduce} for the list tail. Therefore, the list is eventually reduced to a single value by recursively applying {\tt f} from right to left. The first reduce query computes the sum of all elements in $M$ (the list returned by {\tt map}), whereas the second one computes the product of the squares (see second lambda function parameter) of all integers in the range from 1 to 10 that are divisible both by 2 and by 3.

Figure~\ref{fig:highOrder} also includes a high-order function to filter the jsons that satisfy some conditions in one of the fields: {\tt jsFilter}, whose filtering condition parameter is expressed by a function {\tt filterC}, that receives the current json and a field, say with key $K$ and value $V$, to be used for the evaluation. Next, the figure shows two implementations for {\tt filterC}: (1) {\tt select1KV}, which checks whether the json has a field equal to $(K,V)$, and (2) {\tt select1KinV}, which checks whether the value $V$ is included in the list of elements that represents the value for the the field $K$ in the json. Two queries on the variable {\tt emps} defined in Figure~\ref{fig:json} are issued: the first one filter all employees with age 28 and the second one the employees that work for at least the project {\tt "p1"}.

% !TeX spellcheck = en_US

%\begin{lrbox}{\leftbx}%
%	\begin{minipage}{.45\linewidth}%
%		\begin{lstlisting}
%>> ^1+(2+1/3)-(2+1//3);
%1.3333333333333335
%>> ^'A'+1=='B';
%true
%>> x=2+.1;
%>> x*=2;
%>> ^x;
%4.2
%>>x='A'+1=='B';
%>> ^x;
%true
%>> y="Hello "+"Worl"+'d';
%>> ^y;
%Hello World
%>> ^y[0:1]+'i'+y[5:];
%Hi World
%\end{lstlisting}
%	\end{minipage}%
%\end{lrbox}
%
%
%
%
%
%
%
%
%
%\begin{figure}[t]
%	\subfloat[]{\framebox[1.1\width]{\usebox{\leftbx}} }
%
%\end{figure}
%

\section{Imperative Aspects of CalcuList}
\label{sec:imp}

\subsection{Local Variables}

To simplify the writing of a function and, sometime, to optimize its execution (mainly, by avoiding to call twice the same function), it is possible to define {\em local variables} inside a function as follows: {\tt f(...): <LV\mbox{$_1$},..., LV\mbox{$_n$}> expr}. The $n$ variables are only used by the function {\tt f} and are stored into each of the frames of each called {\tt f} instance.

The local variable values are initialized and updated  by means of the so-called {\it Local Setting Commands} (LSC), that are imperative statements without side effects. An LSC is an assignment of an expression value to a local variable, surrounded by adorned curly braces \verb+{!+, \verb+!}+. If {\tt expr} is a simple expression, LSCs can be inserted before it as:  LSC$^*$ \mbox{\tt expr},
where { LSC} is executed before the function starts its execution. Note that, as indicated by the Kleene operator $^*$,  {\it LSC} may indeed consist of a sequence of commands, each of them surrounded by curly braces.

In case {\tt expr} is a conditional expression, the local setting commands may be also inserted before each sub-expression, e.g.,
\vspace{-0.12cm}
 \begin{center}
 	{ LSC$_1^*$ \mbox{\tt cond} \verb+?+ LSC$_2^*$ \mbox{\tt expr}$_1$ \verb|:| LSC$_3^*$ \mbox{\tt expr}$_2$ }
 \end{center}
 \vspace{-0.12cm}

In this case, { LSC$_1$} is executed before the function starts its execution and, on the basis of the result of {\tt  cond}, either (1) { LSC$_2$} is executed before the execution of {\tt expr$_1$} or (2) { LSC$_3$} is executed before the execution of {\tt expr$_2$}. If {\tt expr$_1$} (or {\tt expr$_2$}) is in turn a conditional expression, additional local setting commands can be introduced using the same schema as above.

The usage of LSCs has been inspired by the semantic rules of an Attribute Grammar~\cite{Paakki:1995}. A local variable can be thought of as a inherited grammar attribute and an LSCs as a semantic rule setting the attribute value.

\begin{lrbox}{\leftbx}%
	\begin{minipage}{.89\linewidth}%
		\begin{lstlisting}
>>rotate(L,k) = <n,k1>  {! n=_len(L) !}  k==0? []:  
                {! k1=k%n !} k<0? L[-k1:]+L[:-k1]: L[n-k1:]+L[:n-k1];
>>^rotate([5,1,4,20,15,13], 3);
[ 20, 15, 3, 5, 1, 4 ]
>>^rotate([5,1,4,20,15,13 ],-4);
[ 15, 3, 5, 1, 4, 20 ]
>>part1(x,L,o/2,T0,T1) : L==[]?[T0,T1]: o(L[.],x)? 
      part1(x,L[>],o,[L[.]|T0],T1):  part1(x,L[>],o,T0,[L[.]|T1]);
>>part(x,L,o/2) : part1(x,L,o,[],[]);
>>quicksort(L,o/2) : <T01> L==[]? []: L[>]==[]? [L[.]]: 
  {! T01=part(L[.],L[>],o) !} 
  quicksort(T01[0],o)+[L[.]|quicksort(T01[1],o)];
>>^quicksort([3,11,2,8,6,5], lambda x,y: x<=y);
[ 2, 3, 5, 6, 8, 11 ]

       \end{lstlisting}
\end{minipage}%
\end{lrbox}

%\vspace{-0,3cm}
\begin{figure}
\framebox[1.1\width] {
\usebox{\leftbx}
}  
\caption{\em Functions with Local Variables}
\label{fig:lvar} 
\vspace{-0,7cm}
\end{figure}

As a first example, consider the (clockwise) rotation of the $n$ elements of a list by $k$ positions consists of (i) shifting the elements to the right by $k$ positions and (ii) moving the last $k$ elements to the top of the list. For instance, given the list {\tt L = [ 5, 1, 4, 20, 15, 13 ]}. The rotation of {\tt L} by $k$ positions is {\tt [ 15, 13, 5, 1, 4, 20]} if $k = 2$ or {\tt [ 4, 20, 15, 13, 5, 1 ]} if $k=4$. If $k \geq n$, $k$ is replaced by $k \% n$, i.e., the remainder of dividing $k$ by $n$. If $k=0$ the list remains unchanged. If $k<0$, the rotation is anti-clockwise for it is made by (i) shifting the elements to the left by $k$ positions and (ii) moving the first $k$ elements to the end of the list -- this corresponds to a clockwise rotation by $-k$ positions.
The definition of the function {\tt rotate} is shown in Figure~\ref{fig:lvar} and two local variables ({\tt n} and {\tt k1}) are used to simplify both the writing and the implementation. The built-in function {\verb|_len|} returns the length of a list. Observe that the function does not have side effects.

As second example, we implement Quick Sort, which is a well-known sorting algorithm, which runs in time \cal{O}$(n\: {\tt log}\, n)$ in the average case. Let {\tt x} be the list head and {\tt T} be the list tail. Then we partition {\tt T} into the list {\tt T0} of the elements {\tt y} of {\tt T} such that $\tt y \leq x$ (or larger, depending on the the ordering o) and the list {\tt T1} of all other elements. Then, recursively sort {\tt T0} and {\tt T1} and finally return the concatenation of the two resulting ordered lists, separated by the element $x$. The definition of Quick Sort is shown in Figure~\ref{fig:lvar}. Observe that, given {\tt T}, the function {\tt part} returns {\tt T0} and {\tt T1} as two elements of a list after one scan of the list  {\tt T}. For each stage of  {\tt quicksort}, the function {\tt part} is called once inside the {LSC} and the result is assigned to the local variable {\tt T}.

\subsection{Global Variables inside Functions}

One of the peculiar characteristics of {\it pure} functional programming languages is being stateless, i.e. each  computation can only work on  data which are received  as its input parameters   and on data obtained as outputs from operations invoked during its execution. The use of global variables is then not allowed. While this forces the user to develop, and hence reason, in functional style, there are some situations where the ability to have side-effects may simplify  the developed code or even result in a better computational efficiency. 

A function in CalcuList can be simply declared with side effects by adding \verb+*+ next to its name ({\em star function}) so that the usage of a number of global variables is enabled -- the property of a function to be with or without side effects cannot be changed after its first definition. Global variables to be used by a star function must be explicitly listed in its definition body. Then, they not only are made accessible to the function, but also they may have assigned new values during the function execution, as in an imperative programming language. 
Actually also parameter values can be modified during the execution of a star function. Obviously, a star function cannot be called by a non-star one. 

There are situations (e.g., while importing definitions elaborated in previous sessions) in which a user may accidentally erase the value of a global variable by introducing one with the same name. To mitigate the risk of creating unwished homonyms, CalcuList allows the user to assign a label to variables. For instance the command {\tt L: V1,V2,V3} declares three labeled variables {\tt L.V1, L.V2} and {\tt L.V3} -- they are all initialized to null. After its definition, a labeled variable may have assigned a value by a suitable assignment statement. A label definition can be reissued to declare additional variables with that label and to reset previously-defined labeled variables to null.   

To be used inside a function definition, a global variable must be declared as a labeled variable and its label (say \verb|L|), followed by~\verb+*+, must be listed between brackets as \verb+<L*>+ at the beginning of the function definition. The labeled variables are then addressed inside the function without writing their label. It is possible to list more than one label for a function, which are to be included between brackets together with possible local variables.

\subsection{Functions with side effects}
A star  function may have side effects, that is: (1) labeled global variables can be used as terms inside the expression defining the function, in addition to constants, parameters and function calls and (2) both labeled global variables and parameters can be updated inside a function by means of the so-called {\it global setting commands} (GSC). A $GSC$ is an assignment of an expression value to a labeled global variable or to a function parameter and represents an imperative statement with side effects. The syntax of GSCs is similar to the one of LSCs, described above. 
If {\tt expr} is a simple expression, GSCs can be inserted not only before the expression {\tt expr} but also after it:
\vspace{-0.12cm}
\begin{center}
{ GSC$_1^*$} \mbox{\tt expr} {GSC$_2^*$}
\end{center}
\vspace{-0.12cm}
 \noindent
where { GSC$_1$} is executed before the function starts its execution and  { GSC$_2$} at the end of its execution. Note that  {\it GSC$_1$} and {GSC$_2$} may indeed consist of a sequence of commands, each of then surrounded by curly braces {\tt \{! !\}}.

In case {\tt expr} is a conditional expression, the global setting commands may be also inserted before and after each sub-expression, e.g.,
\vspace{-0.12cm}
 \begin{center}
 	{ GSC$_1^*$} \mbox{\tt cond} \verb+?+ GSC$_2^*$ \mbox{\tt expr}$_1$ {GSC$_3^*$}\verb|:| {GSC$_4^*$} \mbox{\tt expr}$_2$ {GSC$_5^*$}
 \end{center}
 \vspace{-0.2cm}

\begin{lrbox}{\leftbx}%
	\begin{minipage}{.89\linewidth}%
		\begin{lstlisting}
>>MATH: zeroD;  
>>div*(x,y): <MATH*>  {! zeroD=false !} y==0 ? 0 {! zeroD=true !}:x/y; 
>>MATH1: numErr, somma;
>> listDiv1*(x,L): <MATH*, MATH1*,d>  L==[]? []:  
  {!d=div(x,L[.]) !} zeroD? {! numErr+=1 !}  ['*' |listDiv1(x,L[>])]: 
  [d | listDiv1(x,L[>])] {!  somma+=d !};
>>listDiv*(x,L):  <MATH1*> {! numErr=0 !} {! somma=0 !} listDiv1(x,L);
>>^listDiv(4,[ 2, 0, -4, 0, 20 ]);
[ 2.0, '*', -1.0, '*', 0.2 ]
>>^MATH1.numErr;
2
>> ^MATH1.somma;
1.2
>>swap*(L,i,j):  <t> true  {! t=L[i] !} {! L[i]=L[j] !} {! L[j]=t !}; 
>>K=[1,2,3,4];
>>^swap(K,1,3);
true
>>^K;
[1, 4, 3, 2]		
       \end{lstlisting}
\end{minipage}%
\end{lrbox}

%\vspace{-0,3cm}
\begin{figure}
\framebox[1.1\width] {
\usebox{\leftbx}
}  
\caption{\em Global Labeled Variables within Functions}
\label{fig:gvar} 
\vspace{-0,7cm}
\end{figure}

In this case, { GSC$_1$} is executed before the function starts its execution and, on the basis of the result of {\it  cond}, either (1) { GSC$_2$} and { GSC$_3$} are executed before and after the execution of {\tt expr$_1$} or (2) { GSC$_4$} and { GSC$_5$} are executed before and after the execution of {\tt expr$_2$}. If {\tt expr$_1$} (or {\tt expr$_2$}) is in turn a conditional expression, additional local setting commands can be introduced using the same schema as above.

Note that, as for LSCs, also the usage of {GSC}s has been inspired by the semantic rules of an Attribute Grammar. In this case, a global variable can be thought of as either an inherited or a synthesized grammar attribute and a {GSC}s corresponds to a semantic rule setting the attribute value. 
In Figure~\ref{fig:gvar} the  variable {\tt  zeroD} is defined with label {\tt  MATH} and is used in the {\tt  div} function to detect division by zero. Next a function {\tt  listDiv} is defined that, given a number {\tt  x} and a list {\tt  L} of elements, returns the list of all divisions of {\tt  x} by every element in {\tt  L} if the division of {\tt  x} by that element is feasible or 
{\tt '*'} otherwise. The overall implementation makes use of two additional global variables with label {\tt  MATH1}: {\tt numErr}, which counts the number of wrong divisions, and {\tt somma}, which compute the sum of all correct divisions. The local variable {\tt d} is also used to store the division result.

As an example of star function without global variables, consider the function,  {\tt swap} reported in Fig.~\ref{fig:gvar} , that swaps two elements $i$ and $j$ of a list {\tt L}. The function always returns {\tt true} after swapping the two elements. The {\tt swap} function  has side effects for it alters the content of its formal parameter {\tt L} and, therefore, of the global variable {\tt K}, which is passed as argument (actual parameter).
\begin{lrbox}{\leftbx}%
	\begin{minipage}{.89\linewidth}%
		\begin{lstlisting}
>>newVx_1(m,j) : j >=m? []: [0 | newVx_1(m,j+1)];
>>newVx(m) : m <= 0? []: newVx_1(m,0);
>>triangLS_sum(A,X,i,j,n1): j>n1? 0:  
           A[i][j]*X[j]+triangLS_sum(A,X,i,j+1,n1);                                  
>>triangLS_1*(A,B,X,i,n1): i<0? X: A[i][i]==0? exc("matrix A singular"): 
   {! X[i]=(B[i]-triangLS_sum(A,X,i,i+1,n1))/A[i][i] !}  
   triangLS_1(A,B,X,i-1,n1);
>>triangLS*(A,B): <n>{! n=_len(A) !} n!=_len(A[0])||n!=_len(B)? 
        exc("matrix A and vector B are not conformant"): 
        triangLS_1(A,B,newVx(n),n-1,n-1);
>>^triangLS([[1,2,-1],[0,2,4],[0,0,-2]],[-6.5,3.0,-1.5]);
[ -6.5, 3.0, -1.5 ]	
       \end{lstlisting}
\end{minipage}%
\end{lrbox}

%\vspace{-0,3cm}
\begin{figure}
\framebox[1.1\width] {
\usebox{\leftbx}
}  
\caption{\em Resolution of an Upper-Triangular Linear Equation System}
\label{fig:matrix} 
\vspace{-0,7cm}
\end{figure}

An example of star function that shows the capability of adding imperative statements while preserving the basic functional style is presented in Figure~\ref{fig:matrix}: resolution of a linear equation system $A \times X = B$ for the case $A$ is a an upper triangular matrix, i.e., all the entries below the main diagonal are zero. Note that a necessary and sufficient condition to have finite solutions is that the diagonal does not contain zeros. 
Using an imperative language, the resolution can be easily done in quadratic time. (The complexity is measured w.r.t. the size $n$ of the input matrix.) The star function {\tt triangLS} implements a classical imperative algorithm on matrices in a straightforward way; however, its overall complexity is cubic as shown next. The function, after first checking wether the matrix $A$ and the vector $B$ are conformant (i.e., $A$ is quadratic, say with size $n$, and $B$ has size $n$ as well), calls {\tt triangLS\_1} by setting the index $i=n-1$. Then  {\tt triangLS\_1} computes $X[i] = B[i]/A[i][i]$, and recursively continues the computation by decreasing the index $i$ down to 0 so that, at a generic step $i$,  $X[i] = ( B[i] - \sum_{ i < j < n} A[i][j] \cdot X[j] ) / A[i][i]$. Note that if the function finds a zero in the diagonal, it stops the computation by raising an exception. The cost of the summation in the generic step is not linear but quadratic as the direct access is not available in CalcuList to access list elements. The complexity can be improved by rewriting the functions {\tt triangLS\_sum} and {\tt triangLS\_1} as follows:

\vspace{0.1cm}
\noindent
{\tt \footnotesize
triangLS\_sum(Ai,Xi): Xi==[]? 0: Ai[.]*Xi[.]+triangLS\_sum(Ai[>],Xi[>]);  \\                          
triangLS\_1*(A,B,X,i,n1) : <Ai,Xi> i<0? X: \\
\mbox{$\ \ \ \ \ \ \ \ \ \ $} \{! Ai =i==0? A[i]: A[i][>i-1] !\} \{! Xi = i==0? X: X[>i-1] !\}\\
\mbox{$\ \ \ \ \ \ \ \ \ \ \ $}Ai[0]==0? exc("matrix A is singular"): \\
\mbox{$\ \ \ \ \ \ \ \ \ \ $} \{! Xi[0]=(B[i]-triangLS\_sum(Ai[>],Xi[>]))/Ai[0] !\}  \\
\mbox{$\ \ \ \ \ \ \ \ \ \ \ $}triangLS\_1(A,B,X,i-1,n1);
}

\vspace{0.1cm}
\noindent
where $L[>i]$ return a shallow copy (i.e., without actually scanning all elements) of the sublist of $L$ starting from the element $L[i+1]$ up to the end. In this way the summation at the generic step has linear time complexity and, therefore, the overall computation is done in quadratic time. 

%As a small experiment, we randomly generated two problem instances: the first one with $n=50$ and the second one with doubled size, i.e., $n=100$. Using the functions in Figure~\ref{fig:matrix}, the number of clops increased of 6.87 times while the increase was only of 3.80 times for their improved versions. Moreover, the number of clops by the improved version were around 23\% of the number of clops by the first version for $n=50$ and 13\% for $n=100$.

Figure \ref{fig:jsonSE}  includes a star function that manipulates a list of jsons. As a list of jsons can be though of as a documentary database, star functions have the crucial role of updating the database. As an example we have written the function {\tt giveBonus} that assigns a bonus of a given amount to all employees in a list -- the amount is incremented if an employee already has a bonus. Obviously the function has side effects as it updates the elements in the parameter {\tt emps}. We present a query that assigns a bonus of 100 Euro to all employees in the global variable {\tt emps} (defined in Figure \ref{fig:json}) who work in project {\tt p1}.

We conclude by mentioning that versions of GSCs can be used to implement advanced printing features: they are called {\em Global Printing Commands} ($GPS$s) and are enclosed between  the curly braces {\tt \{\verb|^|  \verb|^|\}}. The print output can be redirected to an external tex file by the command {\tt>>({\em file\_name})}. There is also a command {\tt<<({\em file\_name})} for reading a CalcuList value, such as a list or a json, from an external text file.

\begin{lrbox}{\leftbx}%
	\begin{minipage}{.89\linewidth}%
		\begin{lstlisting}
>>giveBonus*(emps,value):emps==[]?true:emps[.]["bonus"]==null? 
    {! emps[0]["bonus"]=value !} giveBonus(emps[>],value):
    {! emps[0]["bonus"]+=value !} giveBonus(emps[>],value);
>>^giveBonus(jsFilter(emps,select1KinV,"projects","p1"),100);
true
>>^emps;
[ { "name": "e1", "age": 30 }, 
  { "name": "e2", "age": 32, "projects": [ "p1", "p2" ], "bonus": 100 }, 
  { "name": "e3", "age": 28, "projects": [ "p1", "p3" ], "bonus": 100 } 
]
\end{lstlisting}
\end{minipage}%
\end{lrbox}

%\vspace{-0,3cm}
\begin{figure}[t]
\framebox[1.1\width] {
\usebox{\leftbx}
}  
\caption{\em Json Manipulation in CalcuList}\label{json}
\label{fig:jsonSE} 
\vspace{-0,7cm}
\end{figure}

\section{Conclusion}
\label{sec:conc}

In this paper we have presented a new educational functional programming language extended with imperative programming features: CalcuList, whose imperative features are enabled under explicit request by the programmer.
CalcuList expressions and functions are first compiled and then executed each time a query is issued. The object code produced by a compilation is a program that will be eventually executed by the CalcuList Virtual Machine (CLVM). The system provides an assembler component  to run CLVM programs using an assembler language.
The  CalcuList programming environment has been implemented as a small-sized Java project in Eclipse 4.4.1 with 6 packages and 20 classes all together. The Jar File for using Calculist may be dowloaded from the link in \cite{CLR4}. The size of this file is rather small: 134 kb. The draft of a tutorial on CalcuList may be dowloaded from the link in \cite{CLTutor}.

CalcuList has been used since 2011 by the first author as a didactic tool for his course of Formal Languages, thought at the first year of the Master in Computer Engineering at University of Calabria. The course used to focus on basic notions of languages, grammars and compilers and on two styles of declarative paradigms: logic programming (mainly Prolog and Datalog)  and a functional language. As it was hard to introduce another language to illustrate the functional paradigm in a limited number of teaching hours, the first idea was to focus on the functional features supported by classical imperative languages. The results were very disappointing: the declarative style was at most adopted as syntactic sugar that flied away at the first serious attempt to replace iteration with recursion. At that point, also considered that students were eager to learn new emerging languages such as Python, the first author invented a new functional language that at the appearance looked as Python, but it did not support any iterative construct. So, to pass the exam, a student had to eventually learn a functional programming language, although with a practical look. 

The first version of CalcuList was rather rudimental and a number of features have been later on added year after year on demand. For instance, when it became popular the MapReduce programming paradigm, higher-order functions were introduced but without the possibility to return a function as this feature was not necessarily required to explain MapReduce. Later on, as the manipulation of jsons became a new frontier, they were added to CalcuList as first class objects. In addition, to enable the update of a json document  while filtering it, Global Setting Commands (GSCs) were introduced to perform operations with side effects using a style borrowed by the semantic rules of attribute grammars. In sum, in six years CalcuList has moved from  a simple pure functional core to a flexible functional programming environment with some (limited and controlled) imperative  features that sometimes may very much simplify a strict pure functional notation. 

We conclude by mentioning that,
stimulated by the accurate and competent remarks by three anonymous referees of an earlier version of this paper, we  started to work on the following extensions of CalcuList:
\vspace{-0.2cm}
\begin{enumerate}
\item infer types for function parameters and returned value, although we shall not be able to provide a complete inference system as we shall  preserve the present weak typing for lists and jsons;
\item provide a statical test of whether a function with shallow list operations has  side effect or not -- the extension is non complex if applied to each function separately from the other ones, but the real challenge  is to consider a group of functions where a function may pass a list parameter according to the typical tail recursion scheme.

\end{enumerate}

\bibliographystyle{abbrv}
\bibliography{cbiblio}

\end{document}